# Two planetary nebulae in the Sagittarius Dwarf Galaxy

Albert A. Zijlstra and J. R. Walsh

European Southern Observatory, Karl-Schwarzschild-strasse 2, D-85748 Garching bei München, Germany

**Summary.** Two planetary nebulae are shown to belong to the Sagittarius Dwarf Galaxy, on the basis of their radial velocities. This is only the second dwarf spheroidal galaxy, after Fornax, found to contain planetary nebulae. Their existence confirms that this galaxy is at least as massive as the Fornax dwarf spheroidal which has a single planetary nebula, and suggests a mass of a few times $10^7 \, M_\odot$. The two planetary nebulae are located along the major axis of the galaxy, near the base of the tidal tail. There is a further candidate, situated at a very large distance along the direction of the tidal tail, for which no velocity measurement is available. The location of the planetary nebulae and globular clusters of the Sagittarius Dwarf Galaxy suggests that a significant fraction of its mass is contained within the tidal tail.

## 1. Introduction

The unexpected discovery of a dwarf spheroidal galaxy in Sagittarius (Ibata et al. 1994, 1995) has shown the elusive nature of these Galactic companions. At present there are 9 such systems known in the Galactic halo, ranging in mass from $2 \times 10^7 \, M_\odot$ (Fornax) down to a few times $10^5 \, M_\odot$ (Ursa Minor). The Sagittarius Dwarf is the closest of these, at a distance of $25 \pm 3$ kpc, and possibly also the most massive. The Galactocentric distance is only 16 kpc and it is being disrupted by the Galaxy into which it will eventually dissolve. It was found accidently but contains several globular clusters which were known before. By far the most luminous of these is M54, which may even be the nucleus of the new dwarf (Ibata et al. 1995; Da Costa & Armandroff 1995; Sarajedini & Layden 1995).

From star counts, Ibata et al. found an angular size for the Sagittarius Dwarf of $10° \times 6°$ corresponding to $6 \times 2.5$ kpc. This extent is indicated in Figure 1. Ibata et al. suggested that it is significantly larger based on its associated globular clusters, and estimate that the outer contours enclose less than 50% of the total mass. Mateo et al. (1996; see also Fahlman et al. 1996) confirmed that the major axis of the galaxy can be traced to at least $b = -25°$, with a remarkably uniform surface brightness between $4°$ and $9.7°$ from M54; they suggest that the Sagittarius Dwarf extends at least $20°$ along its major axis. Alard (1996) at the same time found evidence for a population of RR Lyrae stars on the opposite side, closer to the Galactic plane. Mateo et al. suggested that the enlarged extent is more easily explained if the galaxy has already undergone an earlier encounter with the Galaxy which produced a long tidal tail (Piatek & Pryor 1995) extending on both the leading and trailing side (Moore & Davis 1994). Proper motion studies suggest that the galaxy is moving towards the Galactic plane (Irwin et al. 1995) but this is not yet certain.

The metallicity–luminosity relation for dwarf spheroidals (e.g. Caldwell et al. 1992) suggests that Sagittarius is significantly more luminous and massive than the other Galactic dwarf spheroidals, rivaling the dwarf ellipticals around M31. However, despite its large size, its total mass is similar to that of Fornax. A significant fraction of the galaxy may already have dispersed into the Galactic halo. It is interesting that Sagittarius has four globular clusters (including the highly luminous M54), the same as Fornax (e.g. Minniti et al. 1996). The other, smaller Galactic dwarf spheroidals are not known to contain globular clusters.

Here we argue that two previously known Galactic planetary nebulae also belong to this galaxy. This makes it the second dwarf spheroidal to have planetary nebulae, and indicates that its present mass is indeed comparable to, or slightly larger than that of Fornax.

## 2. Planetary nebulae in the Sagittarius Dwarf Galaxy

Of the approximately 1000 known Galactic planetary nebulae (PN) (Acker et al. 1992), three occur in the direction of the Sagittarius dwarf galaxy. One of these (003.8−17.1, Hb8) is certainly not related, based on its heliocentric radial velocity of $-172 \, \mathrm{km \, s^{-1}}$: the velocity of the Sagittarius dwarf galaxy is $+140 \, \mathrm{km \, s^{-1}}$ with a velocity dispersion of $10 \, \mathrm{km \, s^{-1}}$. The other two PN are 004.8−22.7 (He2-436) and 006.8−19.8 (Wray 16-423) for which no velocities have been published. All three PN are far from the plane of the Galaxy, in a region devoid of other PN in spite of being at longitudes within the area of the bulge. Two further candidates which may possible be related to the Sagittarius Dwarf are discussed below. The candidates are listed in Table 1. Finding charts can be found in Acker et al. (1992).

Radial velocities of He2-436 and Wray 16-423 were measured using optical spectroscopic observations. The [O III] 5007Å line was observed using the 1.4m CAT telescope at La Silla, at a resolving power of 30000 with a $2''$ slit width.

**Figure 1.** Iso-density contours for the Sagittarius Dwarf galaxy as determined by Ibata et al. (1995), with RA and DEC in B1950 coordinates. The galaxy is known to extend significantly further along its major axis (the tidal tail). The crosses indicate three of the four associated globular clusters. The plus signs indicate the two planetary nebulae we suggest belong to this galaxy.

**Table 1.** The planetary nebulae in the direction of the Sagittarius Dwarf galaxy

| PN | name | RA (J2000) | DEC (J2000) | $V_r$(hel) ($\mathrm{km\,s^{-1}}$) | $V_{exp}$ ($\mathrm{km\,s^{-1}}$) | log F(H$\beta$) ($\mathrm{ergs\,cm^{-2}\,s^{-1}}$) |
|---|---|---|---|---|---|---|
| *likely members* | | | | | | |
| 004.8−22.7 | He2−436 | 19 32 07.3 | −34 12 33 | +132.9 | 13 | −12.17 ± 0.03 |
| 006.8−19.8 | Wray16-423 | 19 22 10.6 | −31 30 40 | +133.1 | 15 | −12.0 ± 0.3 |
| *possible members* | | | | | | |
| 0.09.6−10.6 | M3−33 | 18 48 12 | −25 28 50 | +180 | 24 | −12.0 ± 0.2 |
| 006.0−41.9 | PRMG1 | 21 05 53.5 | −37 08 17 | | | −13.3 ± 0.4 |
| *not a member* | | | | | | |
| 003.8−17.1 | Hb8 | 19 05 36.4 | −33 11 39 | −172 | | −11.82 ± 0.03 |

The spectra are shown in Figure 2. A heliocentric velocity of $+132.9\,\mathrm{km\,s^{-1}}$ for He 2-436 and $+133.1\,\mathrm{km\,s^{-1}}$ for Wray 16-423 is derived, with uncertainties of less than $2\,\mathrm{km\,s^{-1}}$. The velocities were derived from a first-moment analysis of the profiles, and corrected for the Earth's motion. The uncertainty is mostly due to the asymmetries in the PN emission-line profiles, which make the result dependent on the method of analysis (e.g. Spyromilio 1995). The estimate of the uncertainty is derived from comparing a larger sample of observations with data available in the literature (in preparation).

The objects are indicated in Figure 1: they are situated below the main body of the galaxy, being about 5° displaced along the major axis and lying at the base of the tidal tail found by Mateo et al. (1996). This is consistent with the distribution of the globular clusters which, with the exception of M54, also trail the galaxy. Ibata et al. (1995) found that the velocity changes by less than $5\,\mathrm{km\,s^{-1}}$ between $b = -12°$ and $b = -20°$ (see also Velazquez & White 1995). The velocities of the two planetary nebulae are in excellent agreement with the expected velocity of the Sagittarius dwarf galaxy at these positions. The velocity gradient is important for orbit determinations (Johnston et al. 1995, Velazquez & White 1995).

For He 2-436, de Oliveira-Abans & Faúndez-Abans (1991) suggested a distance of only 170 pc based on an extinction–distance relation in this direction (which would have made it one of the nearest planetary nebulae). However, their data is unconvincing and, at this Galactic latitude, it is very difficult to establish more than a lower limit to the distance. The extinction towards the Sagittarius Dwarf is low, at $E_{B-V} = 0.15$ and varies by less than 0.05 in $E_{B-V}$ over the central area (Ibata et al. 1995, Mateo et al. 1995).

The expansion velocities listed in Table 1 are calculated as half the total width of the [O III] line profile at the half-power points, corrected for the instrumental resolution and thermal broadening at $T_e = 10^4$ K. The values are typical of nebulae originating from low-mass stars (high-mass stars tend to have somewhat higher expansion velocities). An indication of the luminosity can be derived from the H$\beta$ flux (e.g. Zijlstra & Pottasch 1989). For He 2-436 we derive 1200 L$_\odot$ with an uncertainty of up to a factor of 2. This luminosity also is consistent with a low-mass progenitor star.

**Figure 2** [O III] 5007Å spectra of the two suggested PN in the Sagittarius Dwarf. Panel A shows He 2−436 and panel B Wray 16−423. Intensity is in arbitrary units.

He 2-436 has been detected by IRAS, with flux densities of 0.47 and 0.63 Jy at 12$\mu$m and 25$\mu$m. The IRAS colour is consistent with a fairly young, compact nebula. He 2-426 is known to have a high density in combination with a small angular size (Kingsburgh & Barlow 1992; Webster 1975), consistent with the IRAS detection.

No spectra of He 2-436 and Wray16-423 have been published, although a few values for individual line strengths and line ratios are available in the literature (Tylenda et al. 1994; Kingsburgh & Barlow 1992; Acker et al. 1992). Moderately high temperatures have been proposed for the central stars: $T = 9 \times 10^4$ K and $7.3 \times 10^4$ K for He 2-436 and Wray 16-423, respectively (Preite-Martinez et al. 1989). Abundance determinations have not been derived. It would be particularly interesting to study the nebular abundances for, e.g., oxygen and argon which are not affected by dredge-up and therefore trace the initial abundances of the central star. The angular diameters have not been measured: from the profile along the slit in the CAT data upper limits can be given of 5″. (The diameter given in Acker et al. (1992) for He 2-436 is in fact an upper limit, as mentioned in the errata to the catalogue.)

There are two other possible candidates for planetary nebulae belonging to the Sagittarius Dwarf. The first is 009.6−10.6 (M3-33) for which Schneider et al. (1982) list a velocity of +174 ± 10 km s$^{-1}$. We have measured a velocity of +180±2 km s$^{-1}$ for this object from our CAT data, compared to an expected velocity for the galaxy of +135 km s$^{-1}$ from the calculations of Velazquez & White (1995). The diameter is 5″: the combination of the two make it likely (but do not prove) that 009.6−10.6 is not related to the Sagittarius Dwarf.

The second candidate is 006.0−41.9 (PRMG 1) which is considered part of the Galactic halo. Although located more than 20° from the centre of the Sagittarius Dwarf, it lies closely along the direction of the tidal tail which makes it an interesting candidate. No velocity is known for this nebula, for which an abundance of [O/H]= 8.4 has been reported (Pasquali & Perinotto 1993) corresponding to −0.5 dex compared to solar. However, Pena et al. (1989) suggest an angular diameter of 8″, derived from long-slit spectroscopy. If confirmed, such a large diameter would argue in favour of a foreground object.

## 3. Discussion

Only one planetary nebula was previously known in a Galactic-halo dwarf spheroidal, discovered by Danziger et al. (1978) in Fornax. From the mass and number of PN in the LMC, they argue that one would expect about 0.5 PN in Fornax and a much smaller number in the other dwarf spheroidals.

**Table 2.** Number of PN per solar mass

| Population | ratio |
| --- | --- |
| LMC | $1.9 \times 10^{-7}$ |
| SMC | $3.3 \times 10^{-7}$ |
| Galactic disk | $3 \times 10^{-7}$ |
| Galactic bulge | $1.5 \times 10^{-7}$ |
| Globular clusters | $5 \times 10^{-8}$ |
| Fornax | $5 \times 10^{-8}$ |

To calculate the expected number of PN in dwarf spheroidals, Table 2 lists the total number of PN per solar mass in several environments. The numbers for the LMC, SMC and bulge are taken from Pottasch (1984). For the Galactic disk, a mass of $8 \times 10^{10}$ M$_\odot$ and 23000 PN (Zijlstra & Pottasch 1991) are assumed. The number for the globular clusters is derived from five planetary nebulae known in the Galactic globular clusters (including three reported in Jacoby et al. 1995) and assuming a total mass of $1 \times 10^8$ M$_\odot$ (adding up the individual cluster masses given in Pryor & Meylan 1993 and correcting for the total number of known Galactic globular clusters). The number for Fornax is based on its single planetary nebula and a mass of $2 \times 10^7$ M$_\odot$. In view of the rather different star-formation histories, the ratios are surprisingly close. Taking the value of $0.5 \times 10^{-7}$ PN M$_\odot$$^{-1}$ as representative for the Sagittarius Dwarf, a mass for this galaxy of $\sim 4 \times 10^7$ M$_\odot$ is indicated based on the two confirmed PN. Obviously this is uncertain and the derived mass would double if the two PN listed as possible members in Table 2 would be found to be member of the Sagittarius Dwarf. However, although Sagittarius may be slightly larger than Fornax, it is presently not as massive as the dwarf ellipticals surrounding M31. The posi-tions of the PN and globular clusters of the Sagittarius Dwarf suggest that a significant fraction of its mass is found in the tidal extension.

Could more PN be discovered in Sagittarius? The central areas are close to the bulge and has probably been fully covered in previous surveys. However, along the tidal tail a few objects may have been missed. The completeness of the PN samples at high Galactic latitudes is discussed by Zijlstra & Pottasch (1991). At low Galactic latitude, confusion with the bulge becomes too large. However, there are only a few PN known in this region with the velocities near to that of the Sagittarius Dwarf and they can be explained as bulge members. Thus, although more PN could be discovered in outlying regions, a large PN population in Sagittarius is not expected.

The metallicity of the Sagittarius Dwarf [Fe/H] $\approx -1.1 \pm 0.3$ (Mateo et al. 1995) which is higher than for the other Galactic dwarf spheroidals. There are indications of a minority younger component ($\sim 4\,\mathrm{Gyr}$) with a higher metallicity (see also Sarajedini & Layden 1995). An intermediate-age population is common among dwarf spheroidals, although its size varies greatly between individual galaxies (e.g. Mighell 1990). The abundances derived for the planetary nebula in Fornax (Danziger et al. 1978) establish it as part of the more metal-rich, intermediate-age population. It will be interesting to test whether the same is true for the Sagittarius PN. One of the PN known in Galactic globular clusters, K648 in M15, exists in a very metal-poor environment, showing that not all PN originate from intermediate stellar populations.

The two [O III] profiles shown in Figure 2 are both asymmetric, indicating either somewhat higher densities on the near side of the nebulae or significant internal extinction. Asymmetries can be caused by movement with respect to the interstellar medium, as is seen in the planetary nebula in the globular cluster M22 (Borkowski et al. 1993) but not in K648 (Bianchi et al. 1995). However, such an interaction would be expected in the opposite direction to that observed, based on the velocity vector of the Sagittarius Dwarf galaxy. Also no HI has yet been found in this galaxy (Koribalski et al. 1994). Internal extinction is possible although this is rarely important for planetary nebulae. We note that if asymmetries are seen with high-resolution imaging, e.g. with HST, it could help determining the direction of the proper-motion vector of the Sagittarius Dwarf.

Finally, it should be noted that the association of these two PN with Sagittarius is important for planetary-nebula research. There are very few Galactic PN with accurately known distances. The two best cases are the two PN in globular clusters, but it is not clear whether these are typical planetary nebulae. The extra-galactic nature of He 2-436 and Wray 16-423 results in a well-determined distance, while at the same time they are still conveniently nearby, being members of our nearest satellite. A detailed study of both objects is recommended for this reason, as well as for abundance and kinematic studies of the Sagittarius Dwarf Galaxy.

## 4. Conclusion

The newly discovered dwarf spheroidal in Sagittarius is found to contain two planetary nebulae. This confirms that it is similar in mass to the Fornax dwarf spheroidal galaxy, which contains one PN. The ratio of planetary nebulae in Sagittarius to that of Fornax is similar to the ratio of globular clusters. A present-day mass for Sagittarius of $\sim 4 \times 10^7\,\mathrm{M}_\odot$ is suggested: it is probably not much larger than Fornax. The PN are located near the base of the tidal tail and have velocities consistent with this position inside the galaxy. The two planetary nebulae offer the prospect of direct abundance determinations for the Sagittarius Dwarf galaxy.